\documentclass{PoS}

\title{Summary and Outlook: \\ 2015 Lepton-Photon Symposium}

\ShortTitle{Summary and Outlook}

\author{\speaker{John Ellis}\thanks{Work supported in part by the European Research Council via the Advanced Investigator
Grant 267352 and by the UK STFC via the research grant ST/L000326/1.}\\
        Theoretical Particle Physics and Cosmology Group, Department of Physics, \\
King's College London, Strand, London WC2R 2LS, U.K; \\
Theory Division, Physics Department, CERN, CH 1211 Geneva 23, Switzerland\\
        E-mail: \email{John.Ellis@cern.ch}\\
        ~\\
      {\tt ~~~~~~~~~~KCL-PH-TH/2015-41, LCTS/2015-28, CERN-PH-TH/2015-225}  }

\abstract{What life is there after Higgs? Specifically, what physics lies beyond the Standard Model (SM)?
These are the biggest questions in particle physics today, and my talk is oriented towards efforts to answer them at the
LHC and elsewhere.}

\FullConference{International Symposium on Lepton Photon Interactions at High Energies\\
		 17-22 August 2015\\
		 University of Ljubljana, Slovenia}

\begin{document}

\section{Introduction}

\begin{figure}[htb]
\centering
\includegraphics[height=3in]{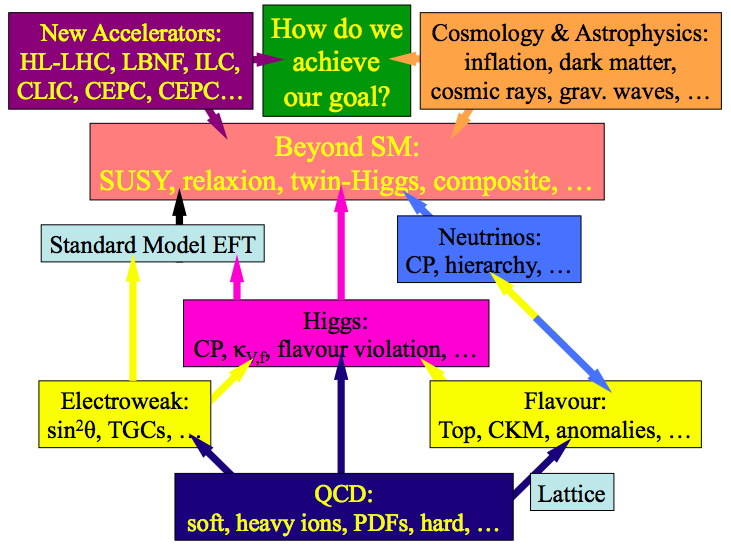}
\caption{\it Our search for physics beyond the SM is rooted in our understanding of the SM,
and will require input from astroparticle and accelerator experiments, as well as define their
future agendas.}
\label{fig:outline}
\end{figure}

Fig.~\ref{fig:outline} outlines the plan of this talk. There are many reasons to think that there must be physics
beyond the SM, including the apparent instability of the electroweak vacuum, the existence
of dark matter, the origin of the cosmological matter-antimatter asymmetry, the masses of
neutrinos, the hierarchy of mass scales in physics, cosmological inflation and the need for a
quantum theory of gravity. The long-term future of particle physics
hinges upon the nature of whatever physics lies beyond the SM, and how we can discover
and study it. As discussed in successive Sections of this talk,
our quest for this new physics must start from our knowledge of
the SM and be advised by theoretical ideas what may lie beyond it. As well as the LHC,
our tools for discovering and studying it include high-intensity low-energy
experiments as well as astroparticle physics and future experiments at the high-energy frontier.

\section{QCD}

Understanding QCD is the essential basis for physics at the LHC, including the search for
any physics beyond the SM. QCD provides backgrounds, in the forms
of the underlying events, jets and pile-up, it provides tests of the SM via multi-jet, $W$, and $Z$ production,
it is crucial for unravelling top physics, notably for determining the top mass, and precision QCD
calculations of hard processes are essential for calculating the production of new particles such as the Higgs boson,
distinguishing them from backgrounds and studying them quantitatively.

However, most QCD processes at the LHC are soft: they cannot, in general, be calculated accurately from
first principles, and cosmic-ray studies as well as QCD physics need them to be 
modelled accurately~\cite{Grosse}. Run I of the LHC
has already helped cosmic-ray physicists~\cite{Perrone} by measuring and reducing the extrapolation uncertainties in the total
and inelastic $pp$ cross-sections. The improved modelling of soft QCD has reduced the uncertainty in
the nuclear composition of ultra-high-energy cosmic rays, and additional help could be provided by measuring
$p$-Oxygen collisions at the LHC. The first LHC Run-II paper contained measurements of the multiplicity and
pseudo-rapidity distribution at 13 TeV~\cite{CMSeta,CMS13}, which provide further discrimination between models: currently,
EPOS-LHC seems to provide the best available description.

\begin{figure}[htb]
\centering
\includegraphics[height=2in]{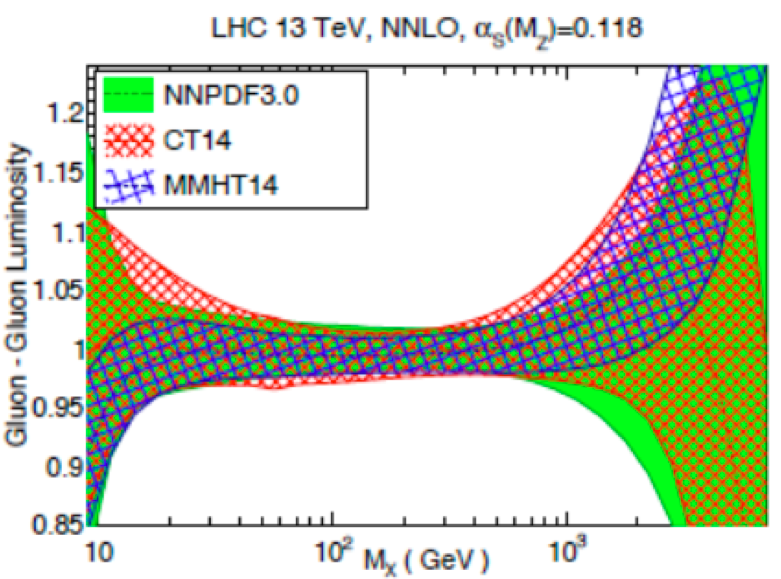}
\includegraphics[height=1.9in]{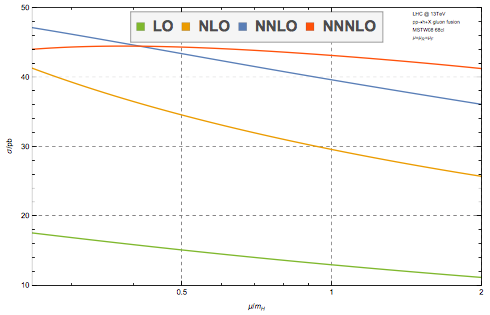}
\caption{\it Left panel: Recent determinations of the gluon-gluon luminosity function for Higgs production at
13~TeV agree to within ${\cal O}(2)$\%~\protect\cite{deVisscher,Zanderighi}. Right panel: The renormalization-scale
dependence of the NNNLO cross-section for $gg \to H$ leads to an
estimated theoretical uncertainty in  is ${\cal O}(3)$\%~\protect\cite{ADDHM}.}
\label{fig:Higgssigma}
\end{figure}

Parton distribution functions (PDFs) provide the bridge between soft and hard QCD, and this Symposium
witnessed significant reductions in their spread and uncertainties~\cite{deVisscher,Zanderighi}. In particular, the uncertainty in the
gluon-gluon luminosity function relevant for the dominant Higgs production mechanism is now ${\cal O}(2)$\%,
thanks in particular to input from HERA, as seen in the left panel of Fig.~\ref{fig:Higgssigma}. This uncertainty is matched by
the uncertainty in the $gg \to H$ subprocess cross-section, which has now been calculated to N$^3$LO
and has an uncertainty ${\cal O}(3)$\%~\cite{ADDHM}. The resultant improvement in the Higgs production 
cross-section will enable a new era in precision Higgs physics, improving the sensitivity to new physics
beyond the SM. Many other processes, including kinematic distributions for
Higgs production, have been calculated to NNLO, and some NLO electroweak 
calculations are also available~\cite{Zanderighi,Dawson}, making
possible other measurements of Higgs couplings.

Perturbative QCD calculations of jet cross-sections agree with the data over many orders of magnitude,
and make possible an interesting measurement of $\alpha_s$ that is consistent with the world average.
Moreover, they show that $\alpha_s$ runs beyond the TeV scale: into a GUT? Many other cross-sections
for the production of $W$, $Z$ bosons and ${\bar t} t$ pairs with varying numbers of jets, as well as
subdominant Higgs production processes, also show excellent agreement between theory and experiment.

One of the key issues in QCD is the determination of the top mass, which is fundamental in itself but also
vital for our understanding of the stability of the electroweak vacuum, as discussed later. 
The LHC and Tevatron experiments have announced a world average value~\cite{WAmt}:
\begin{equation}
m_t \; = \; 173.34 \pm 0.76~{\rm GeV} \, ,
\label{mt}
\end{equation}
representing a determination at the ${\cal O}(1/2)$\% level. But what have they determined, and how is
it related to the short-distance mass in the SM Lagrangian? The good news is that the relationship between
the pole definition of $m_t$ and the running mass is well under control at NNNNLO~\cite{Melnikov,MSSS}:
\begin{eqnarray}
m_t^{rm pole} & = & m_t^{\rm running}(1 + 0.4244 \alpha_s + 0.8345 \alpha_s^2 + 2.375 \alpha_s^3
+ (8.49 \pm 0.25) \alpha_s^4 + \dots)\, , \nonumber \\
& = & m_t^{\rm running} + (7.557 + 1.617 + 0.501 +(0.195 \pm 0.005) + \dots)~{\rm GeV} \, ,
\label{mtchange}
\end{eqnarray}
where the $(\dots)$ represent uncalculated higher-order terms. However, the
relationship between the pole mass and the definition of $m_t$ used in Monte Carlo codes is less clear:
non-perturbative effects may be important, though no obvious bias has been seen so far. At the moment,
the experimental uncertainty in (\ref{mt}) may be dominant, but new measurements have been reported~\cite{Meyer}:
\begin{eqnarray}
m_t \; =
& 174.98 \pm 0.58 \pm 0.49~{\rm GeV} & \; \; \; \; [D0] \, ,\\
& 172.99 \pm 0.91~{\rm GeV} \quad \quad \; \; \; & [ATLAS] \, , \\
& 172.44 \pm 0.49~{\rm GeV} \quad \quad \; \; \; & \; \; [CMS~\cite{CMSmt}] \, ,
\end{eqnarray}
and the overall error in $m_t$ in (\ref{mt}) may be reduced by ${\cal O}(2)$ during LHC Run II,
in which case better understanding will be needed. 

\begin{figure}[htb]
\centering
\includegraphics[height=2in]{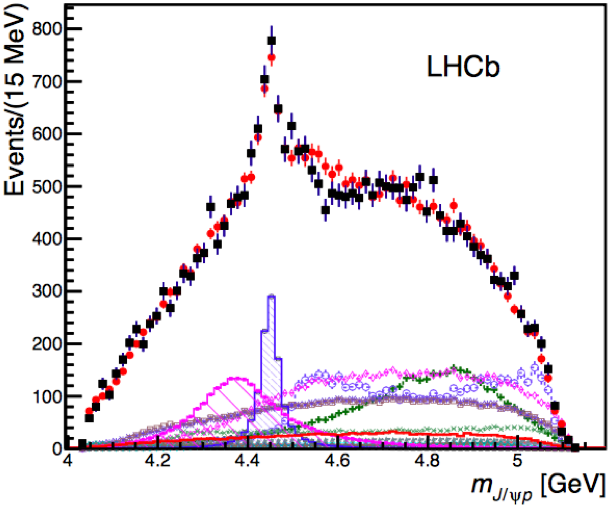}
\includegraphics[height=2.05in]{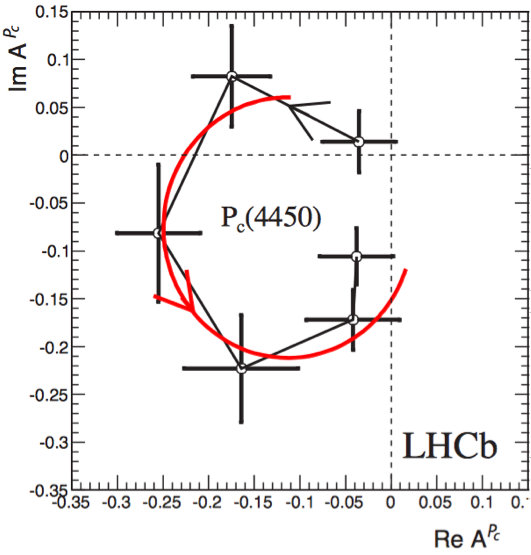}
\caption{\it Left panel: Fitting the LHCb $m_{J/psi p}$ mass distribution requires two exotic states
at 4380~MeV and 4450~MeV~\protect\cite{Skwarnicki}. Right panel: The Argand diagram for the 4450~MeV state exhibits
the behaviour expected for a Breit-Wigner resonance~\protect\cite{Skwarnicki}.}
\label{fig:penta}
\end{figure}

To pentaquark or not to pentaquark? That is the question raised by the discovery of two states
with ${\bar c} c u u d$ quantum numbers reported recently by LHCb~\cite{Skwarnicki}, one of which has a very
nice resonance-like Argand diagram, as seen in Fig.~\ref{fig:penta}. But are they true pentaquark states in a single `bag', or
are they loosely-bound `di-bag' `molecular' states? The latter interpretation may be supported by
fact that the 4450~MeV state has a small decay width, despite having ${\cal O}(400)$~MeV of
phase space to decay into, and the fact that its mass is barely below $m_{\bar \Sigma_c} + m_{D^*}$~\cite{KR},
just as there are several `tetraquark' states close to charm meson-antimeson thresholds~\cite{Bondar}. If it is a 
`molecule', many more `pentaquarks' may be around the corner, and their structure may soon be 
elucidated. Estimates suggest that they could be produced plentifully and observable in photoproduction
experiments at JLAB, in particular~\cite{photo}.

Relativistic heavy-ion collisions are not directly relevant to searches for new physics beyond the SM,
but they may provide an arena for
probing predictions obtained using the AdS/CFT correspondence originating from string theory~\cite{Vuorinen}.
Lattice calculations indicate that there is no phase transition between hadronic and quark-gluon
matter, but a crossover at $T_c = 155 \pm 5$~MeV~\cite{Ukawa}, which is consistent with measurements at
RHIC and LHC~\cite{Arnaldi}. It is expected that $T_c$ should decrease  as the baryon chemical potential $\mu_B$
increases, and that a phase boundary should appear at a critical end-point at some $\mu_B \ne 0$.
Its location cannot yet be calculated reliably, but many experiments (RHIC, CERN SPS, NICA, FAIR)
are underway or planned to search for it. Indications are that the medium beyond the crossover is
perhaps the most perfect fluid known~\cite{vuorinen}. Evidence for this is provided by measurements of azimuthal
anisotropies $v_n$ that can be explained by hydrodynamic flow with a very low viscosity-to entropy
ratio $\eta/s < 0.2$. This very close to the lower bound $\eta/s = 1/4 \pi$ from the stringy AdS/CFT
correspondence~\cite{Son}, which may be useful for calculating other properties of this fluid.

This stringy hydrodynamic picture works very well for azimuthal anisotropies in relativistic ion-ion collisions, and
an analogous (?) collective near-side `ridge' is also seen in $p$-Lead collisions. But it came as a surprise
when CMS observed an analogous near-side ridge in high-multiplicity $pp$ collisions at 8~TeV~\cite{CMSridge}, which
has now been seen also by ATLAS at 13~TeV~\cite{ATLASridge,ATLAS13}. Does this mean that hydrodynamics applies also to
$pp$ collisions? If so, are AdS/CFT ideas applicable? Is it evidence for a colour-glass condensate~\cite{DTV}?
It is to be hoped that the detailed studies possible during LHC Run~II will provide answers.

Many other collective effects have been seen in relativistic heavy-ion collisions, including the
suppression of particle production, jet quenching (which is not seen in $p$-Lead collisions),
$\Upsilon$ suppression, as well as $J/\Psi$ suppression at RHIC and regeneration at the LHC~\cite{Arnaldi}.
They exhibit a rich landscape that we are only beginning to explore~\cite{Vuorinen}.

\section{Flavour Physics}

\begin{figure}[htb]
\centering
\includegraphics[height=2.3in]{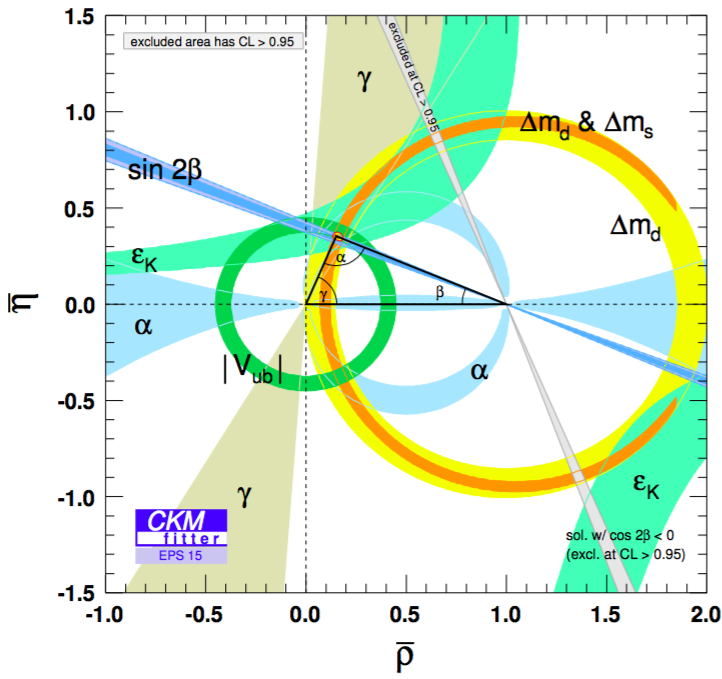}
\includegraphics[height=2.3in]{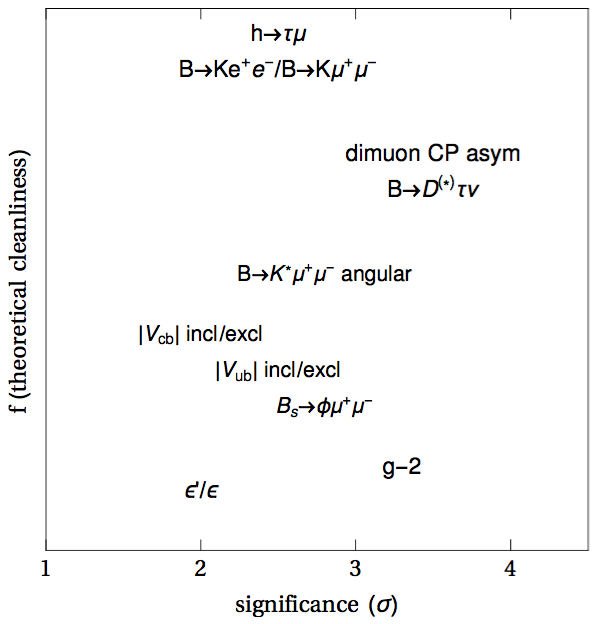}
\caption{\it Flavour physics data are generally consistent with Cabibbo-Kobayashi-Maskawa 
predictions (left panel), but there are a number of apparent anomalies
(right panel)~\protect\cite{Ligeti,Miyabayashi}.}
\label{fig:Flavour}
\end{figure}

Flavour measurements are generally in good agreement with the Cabibbo-Kobayashi-Maskawa
(CKM) description in the SM~\cite{Miyabayashi,Ligeti},
and the unitarity triangle continues to be alive and well in 2015 (with
crucial input from many lattice calculations~\cite{DeTar}), 
as seen in the left panel of Fig.~\ref{fig:Flavour}. The CKM picture has made many successful
predictions, e.g., for many modes of CP violation in $B$ meson decays, and also predicted
successfully (unfortunately) that CP violation in the charm sector would be below the present
experimental sensitivity. However, the data still allow substantial contributions to $B$
physics from effects beyond the SM~\cite{Ligeti}, and its still seems to me an open question whether all TeV physics 
must copy CKM slavishly as in models with minimal flavour violation.

Consider for example $B_{s,d} \to \mu^+ \mu^-$ decay~\cite{Lanfranchi}. The discovery of $B_{s} \to \mu^+ \mu^-$ decay
by LHCb and CMS at the predicted level was a tremendous success for CKM~\cite{CMSLHCb}. However, these
experiments also provide a hint that $B_{d} \to \mu^+ \mu^-$ decay may occur at a level higher
than predicted by CKM, which would require non-minimal flavour violation. 

There are several other anomalies in flavour physics, summarized in the right panel of Fig.~\ref{fig:Flavour},
which one may or may not take seriously.
One of the most intriguing is a hint for $H \to \tau \mu$ decay~\cite{CMSHmutau}, of which more later.
Another is an apparent violation of $e \mu$ universality in $B \to K \ell^+ \ell^-$ decay~\cite{Ligeti}: this would
be such a surprise that my attitude is `wait and see', and the same remark applies to the reported
evidence for a violation of lepton universality in $B \to D^{(*)} \tau \nu$ decay. Concerning the
apparent anomaly in the $P_5^\prime$ angular distribution in $B \to K^* \mu^+ \mu^-$ decay~\cite{Ligeti},
recent theoretical calculations~\cite{Paul} indicate that the QCD uncertainties may have been underestimated
previously. For the same reason, I propose to `wait and see' whether the apparent excess in
the $q^2$ distribution for $B \to \phi \mu^+ \mu^-$ decay will survive. I also note that the value
of the dimuon $A_{FB}$ asymmetry measured at the Tevatron no longer seems to be in
serious conflict~\cite{Ligeti} with SM expectations~\cite{CFM}, I would also take a relaxed attitude towards the
possibility of a violation of CKM unitarity until the discrepancies between the inclusive and
exclusive determinations of $|V_{ub, cb}|$~\cite{DeTar} are ironed out, and I am not losing any sleep
over $\epsilon^\prime/\epsilon$, which is notoriously tricky to calculate accurately~\cite{BGJJ}.

In addition to $B$ physics, there are other interesting experimental probes of flavour physics.
These include measurements of $K \to \pi {\bar \nu} \nu$ decay in both the charged and neutral
channels~\cite{Tschirhart}. The SM makes quite specific predictions for the branching ratios for these decays,
which can be explored in forthcoming experiments. There is currently significant scope for new
physics beyond the SM, particularly in the neutral mode.

Another exciting area is that of the electric dipole moments (EDMs), which violate CP. There has recently been significant
progress by the ACME group in constraining the electron EDM: $d_e < 8.7 \times 10^{-29}$~e-cm~\cite{ACME},
which is an important constraint on many new physics models~\cite{Tschirhart}. However, its significance for
supersymmetry should not be overstated: the CP-violating phases that enter one-loop 
supersymmetric diagrams contributing to EDMs have completely different origins from the Kobayashi-Maskawa phase,
and there are possibilities of cancellations between the phases contributing to the EDMs that are
constrained by current experiments, which could leave open measurable possibilities for other quantities~\cite{cancel}.

An open possibility is that of charged-lepton-flavour violation (CLFV)~\cite{Miller}. This is to be expected given the
observation of neutrino oscillations, but would be unobservably small in the SM. However, it
could be observable in some extensions of the SM, such as supersymmetric models with neutrino masses,
so it is good news that MEG plans an upgrade of its search for $\mu \to e \gamma$, and that renewed
searches for $\mu \to eee$ and $\mu N \to e N$ are also planned. In parallel, flavour factories offer
opportunities to improve the sensitivities to CLFV in $\tau$ decays.

Before leaving this Section, I would like to mention two other searches for new physics in low-energy
physics: $g_\mu - 2$ and the charge radius of the proton~\cite{Tschirhart}. Over a decade after the emergence of a
substantial discrepancy between the experimental measurement of $g_\mu -2$ and SM calculations
based on low-energy $e^+ e^-$ annihilation and $\tau$ decay data, we are no closer to a resolution.
The good news is a new $g_\mu - 2$ experiment is taking shape at Fermilab~\cite{FNALg-2}, that there are also
possibilities at JPARC, and that new $e^+ e^-$
data offer the prospect of substantial reduction in the uncertainty in the SM calculation. As discussed
later, if it holds up the $g_\mu - 2$ discrepancy could be a harbinger of new physics at the TeV scale,
such as supersymmetry. On the other hand, it is difficult for me to imagine a plausible new physics
scenario for the anomaly in the proton charge radius~\cite{Carlson}.

\section{Electroweak Physics at the Tevatron and LHC}

The `gold standard' for electroweak measurements has been provided for over a decade by
LEP and the SLC~\cite{LEPEWWG}. However, the forward-backward asymmetry in $Z$ production at the Tevatron
already gives precision in the electroweak mixing angle $\sin^2 \theta$ that is in the same ball-park 
as individual LEP and SLC measurements, and the Tevatron experiments could do better by combining
all the CDF and D0 results using both $e^+e^-$ and $\mu^+ \mu^-$ measurements~\cite{Einsweiler}. The LHC
experiments have entered the precision electroweak game, but still have much larger
uncertainties, and it will be a long-term challenge to make a competitive determination of $\sin^2 \theta$.

Another class of electroweak measurements is provided by (anomalous) triple gauge couplings (TGCs),
where LEP has led the way so far. LHC measurements of $W^+ W^-, W \gamma$ and $W Z$
final states in Run~I at 8~TeV are already providing competitive constraints on $\Delta \kappa_\gamma, 
\lambda_\gamma, \Delta \kappa_Z, \lambda_Z$ and $\Delta g_1^Z$, LHC measurements of
$Z \gamma$ and $ZZ$ final states constrain other TGCs significantly better than previous measurements
and is starting to constrain quartic gauge couplings~\cite{Einsweiler},
and Run~II will provide significant improvements in the sensitivities to all TGCs. In addition to their
intrinsic importance, the TGC constraints are important ingredients, together with precision electroweak
data and Higgs measurements, in constraining effective field theories (EFTs) of possible physics
beyond the SM, as discussed later.

\section{Higgs Measurements}

The Higgs boson may be the most direct portal to physics beyond the SM.

\begin{figure}[htb]
\centering
\includegraphics[height=2.3in]{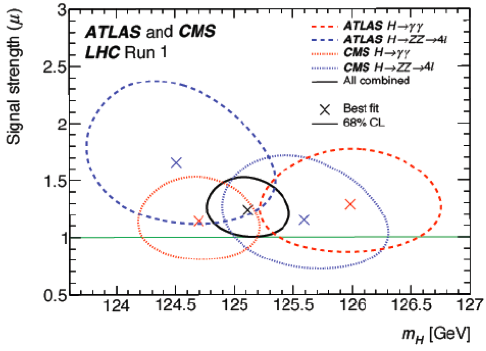}
\includegraphics[height=2.3in]{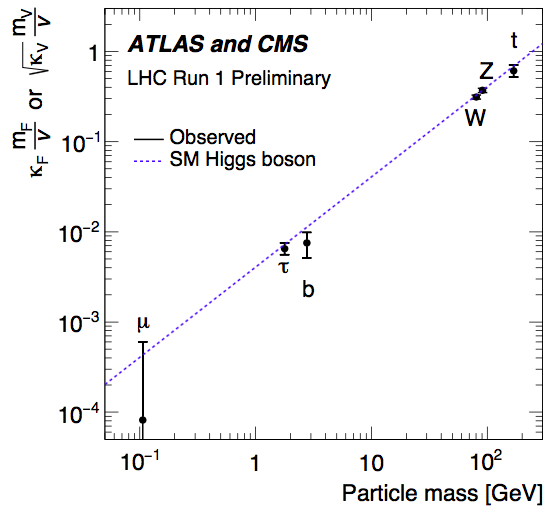}
\caption{\it Left panel: The combination of the ATLAS and CMS measurements of $m_H$~\protect\cite{mH,Farrington}. 
Right panel: Test by ATLAS and CMS of the expected linear dependence of the Higgs couplings to
other particles on their masses~\protect\cite{combined}.}
\label{fig:Higgs}
\end{figure}

The most accurate measurements of the Higgs mass are in the $Z Z^* \to 4 \ell^\pm$ and $\gamma \gamma$
final states. ATLAS and CMS have now combined their results~\cite{mH,Farrington}, as seen in the left panel of Fig.~\ref{fig:Higgs}
to yield a combined value
\begin{equation}
m_H \; = \; 125.09 \pm 0.21~{\rm stat.} \pm 0.11~{\rm syst.} \, .
\label{mH}
\end{equation}
It is noteworthy that statistical uncertainties still dominate, and one can expect that Run~II
measurements will bring the overall uncertainty down to ${\cal O}(100)$~MeV. In addition to
its intrinsic importance, accuracy in $m_H$ enables precision measurements of couplings,
since some Higgs decay rates vary rapidly as functions of $m_H$. Also, like $m_t$, reducing the
uncertainty in $m_H$ is crucial for assessing the stability of the electroweak vacuum
within the SM, as discussed later.

The ATLAS and CMS Collaborations have now also released a combination of their results
on Higgs couplings~\cite{combined}. In particular, they have checked the extent to which the Higgs couplings to
massive particles are proportional to their masses, via parameterizations of the forms
$\kappa_F mf/v$ for fermions and $\sqrt{\kappa_V} m_V/v$ for the $W^\pm$ and $Z^0$,
where $\kappa_F = \kappa_V = 1$ in the SM. As seen in the right panel of Fig.~\ref{fig:Higgs}, their
combined data are currently consistent with the expected linear mass dependence, with the
normalization $v = 246$~GeV expected in the SM.

Both ATLAS and CMS have also made extensive checks of the spin-parity of the Higgs boson
using $Z Z^*, W W^*$ and $\gamma \gamma$ final states~\cite{Farrington}. Alternatives to the expected $0^+$
spin-parity assignment are excluded at $> 99.9$\% CL. Combined CDF and D0 data from the Tevatron
also exclude pure $0^-$ and $2^+$ spin-parity assignments, if the same signal strength as in the SM
is assumed~\cite{TevatronSpin}.

\begin{figure}[htb]
\centering
\includegraphics[height=2.5in]{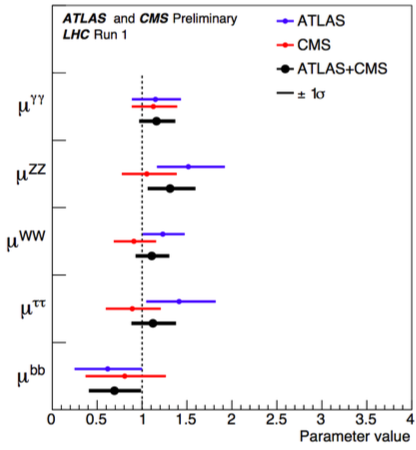}
\includegraphics[height=2.55in]{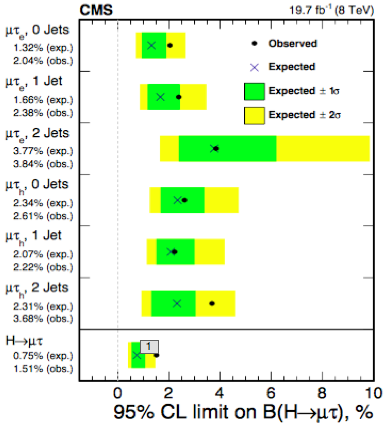}
\caption{\it Left panel: The combination of the ATLAS and CMS measurements 
of the Higgs signal strengths $\mu$~\protect\cite{Farrington,combined}. 
Right panel: Measurement by CMS of the $H \to \mu \tau$ branching ratio~\protect\cite{CMSHmutau}.}
\label{fig:mu}
\end{figure}

The signal strengths $\mu_i$ relative to the SM predictions measured by ATLAS and CMS in 
various channels are shown in the left panel of Fig.~\ref{fig:mu}~\cite{Farrington}, with the overall results
\begin{eqnarray}
\mu \; =
& \; \; 1.18^{+ 0.15}_{- 0.14} \; \; \; & \; \; \; [ATLAS] \, ,\\
& \; \; \; \; \; \; 1.00 \pm 0.14 \; \; \; & \; \; \; \; \; [CMS] \, ,\\
& 1.09^{+ 0.11}_{- 0.10} &  [Combined~\cite{combined}] \, ,
\label{mu}
\end{eqnarray}
which are clearly completely compatible with the SM at the $10$\% level. The dominant sources of uncertainties in the combination
are experimental statistics and the theoretical uncertainty in the signal calculations. Run~II will provide
much increased statistics and a substantial reduction in the signal strength uncertainty is already in hand,
so a reduction in the combined error by a factor $\sim 2$ seems attainable with Run~II.

A fly in this SM Higgs ointment may be provided by flavour-changing Higgs couplings. These
would be very small in the SM or its minimal supersymmetric extension (the MSSM) but all one
can say model-independently is that Higgs exchange should not violate upper bounds on low-energy
flavour-changing interactions. These exclude observable quark-flavour-violating $H$ decays, but
lepton-flavour-violating $H$ decays could be large: $H \to$ either $\mu \tau$ or $e \tau$ could have
a branching ratio ${\cal B} = {\cal O}(10)$\%, though ${\cal B}(H \to e \mu)$ must be $< 2 \times 10^{-5}$~\cite{BEI,Zupan}.
First CMS (see the right panel of Fig.~\ref{fig:mu}~\cite{CMSHmutau}) 
and now ATLAS~\cite{ATLASHmutau} have reported a constraint on ${\cal B}(H \to \mu \tau)$:
\begin{eqnarray}
{\cal B}(H \to \mu \tau) \; =
& 0.84^{+ 0.39}_{- 0.37}\% \; \; \; &  \; \; \; [CMS] \, ,\\
& \; \; \; \; 0.77 \pm 0.63\% \; \; \; & \; [ATLAS] \, .
\label{mutau}
\end{eqnarray}
It would be premature to take seriously this hint at the level of $\sim 2.5 \sigma$,
which could only be accommodated by dint of a surprising complication of the SM
Higgs sector, but let us wait and see. Meanwhile, CMS has also established upper
limits ${\cal B}(H \to e \tau)$\% and ${\cal B}(H \to e \mu) < 0.036$\%~\cite{CMSDPF}, which do not
challenge our prejudices~\cite{BEI,Zupan}.

What do these measurements tell us about the nature of the Higgs boson, and
specifically whether it is elementary or composite? Generally speaking, there have
been two main schools of thought. If it is elementary, the systematic way to tame quadratic
divergences in loop corrections to its mass is to postulate supersymmetry, in which case
the residual corrections to its couplings are expected to be perturbatively small.
Alternatively, in a composite Higgs scenario based on some new strong interactions, one
might expect the corrections to the SM predictions for Higgs couplings to be larger, even if 
one postulates some approximate higher symmetry with the Higgs interpreted as a 
pseudo-Nambu-Goldstone boson (PNGB).

\begin{figure}[htb]
\centering
\includegraphics[height=2.5in]{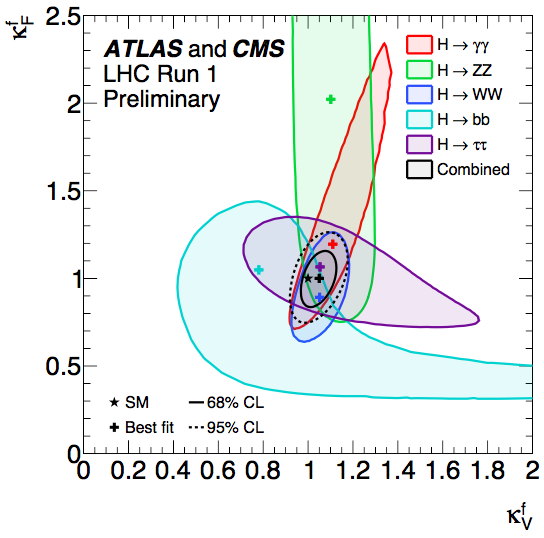}
\includegraphics[height=2.5in]{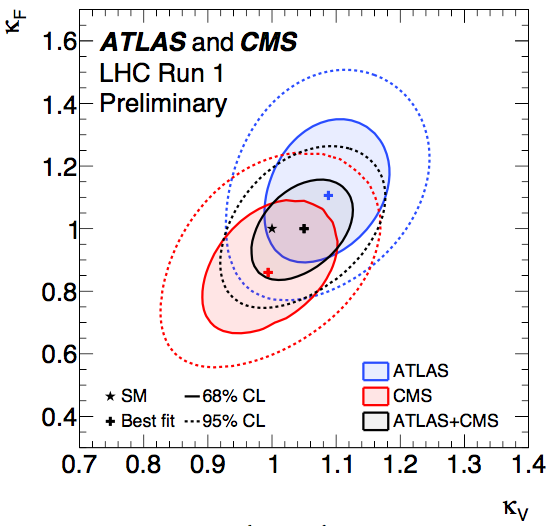}
\caption{\it Left panel: The combination of the ATLAS and CMS measurements of $(\kappa_V, \kappa_F)$
in each measured channel~\protect\cite{combined}. Right panel: The combination of the ATLAS and CMS global fits to 
$(\kappa_V, \kappa_F)$~\protect\cite{combined}.}
\label{fig:kappas}
\end{figure}

One convenient way to analyze this and other models containing a Higgs-like particle is to
introduce factors $\kappa_{V, F}$ modifying its couplings to bosons and fermions,
normalized to unity in the SM~\cite{HXSWG}. Whereas some production and decay modes are
insensitive to the signs of $\kappa_{V, F}$, e.g., $HW$ associated production or
$H \to \tau^+ \tau^-$, others are sensitive to their relative signs, e.g., $H \to \gamma \gamma$ 
or $t H$ associated production. As could be expected from the results (\ref{mu}), the
combined global analysis by ATLAS and CMS is consistent with the SM prediction
$(\kappa_V, \kappa_F) = (1, 1)$, as seen in Fig.~\ref{fig:kappas},
and the negative relative sign is excluded by almost 5$\sigma$~\cite{combined}.
Moreover, there is good consistency between these direct constraints on the Higgs couplings
and indirect constraints coming from precision electroweak data~\cite{Dawson,Ciuchini}. They imply that the parameters
of simple PNGB models must be tuned to resemble the SM at the ${\cal O}(10)$\% level~\cite{Farrington},
as well also as avoiding the prediction of heavier Higgs-like bosons~\cite{Vidal}.

An alternative approach to searching for possible deviations from the SM is to assume
the SM couplings of all particles including the Higgs boson at the level of the renormalizable
terms in the Lagrangian, but allow for for higher-dimensional operators combining these fields
as might appear as relics of higher-energy physics, the SM effective field theory (EFT) approach~\cite{Dawson,Ciuchini}:
\begin{equation}
{\cal L} \; = \; {\cal L}_{\rm SM} + {\cal L}_{\rm eff}: \quad {\cal L}_{\rm eff} \; = \; \sum_n \frac{f_n}{\Lambda^2} {\cal O}_n^{(6)} + \dots
\label{Leff}
\end{equation}
where the ${\cal O}_n^{(6)}$ are a complete set of dimension-6 operators, and the $\dots$ represent terms of
dimension 8 and beyond. There are many such operators, 2499 in the general case, reducing to 59 if
one assumes flavour conservation and family universality, of which 10 are purely bosonic,
but there are also many ways to constrain their coefficients, including precision electroweak data,
TGCs and kinematic distributions in Higgs production and decay, as well as the magnitudes of Higgs couplings.

One example of an analysis where the effects of dimension-6 operators may be important is in the
interpretation of off-shell Higgs production and the constraint it imposes on the total Higgs width, $\Gamma_H$~\cite{Dawson}.
If the off-shell Higgs couplings are the same as those on-shell, ATLAS and CMS data impose upper limits
$\Gamma_H < (5.4, 5.5) \times \Gamma_H|_{\rm SM}$. However, certain of the higher-dimensional operators
in the SM EFT can modify significantly the distributions for $WW$ and $ZZ$ final states, and a model-independent
approach to bounding $\Gamma_H$ would require making a  global fit including their coefficients~\cite{AGPS}.

\begin{figure}[htb]
\centering
\includegraphics[height=2.5in]{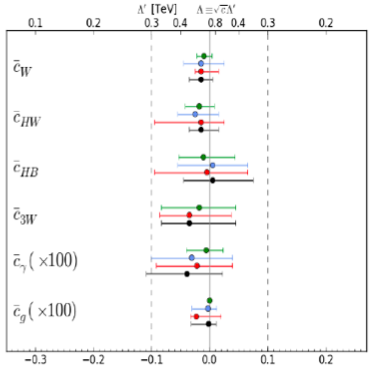}
\includegraphics[height=2.4in]{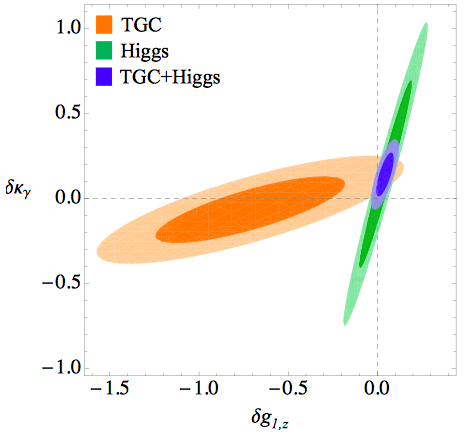}
\caption{\it Left panel: Results from a global fit to the coefficients of dimension-6 operators in the SM EFT: the green
bars are for each operator individually, and the other bars were obtained including all operators, using Higgs
production data (blue), TGCs (red) and all data (black)~\protect\cite{ESY}. 
Right panel: Comparison of the constraints on anomalous
TGCs from TGC data alone (orange) and including Higgs data (blue)~\protect\cite{FGGM}.}
\label{fig:SMEFT}
\end{figure}

Results from one global fit to SM EFT coefficients including precision electroweak data,
the kinematics of associated $HV$ production and LHC TGC measurements as well as Higgs coupling
measurements are shown in the left panel of Fig.~\ref{fig:SMEFT}~\cite{ESY}, and results for constraints on TGCs from
including Higgs data are shown in the right panel of Fig.~\ref{fig:SMEFT}~\cite{FGGM}. The latter illustrates nicely the
complementarity between TGC and Higgs data.

The SM EFT is the most powerful tool to parameterize and search for possible physics beyond the
SM in the electroweak and Higgs sectors, and the LHC Higgs Cross-Section Working Group is
developing recommendations for how to apply it in a standard way.

\section{Escaping the Bondage of the SM}

To paraphrase James Bond, ``The Standard Model is not enough''. There are many concrete physical arguments for this,
of which I list (inevitably) just 007. 1) The electroweak vacuum is not stable in the SM, if one take at face value the
measurements of $m_t$ and $m_H$ and extrapolates naively to high scales without introducing new physics.
2) The SM has no candidate for cosmological dark matter. 3) The SM does not explain the origin of matter.
4) The SM does not include the mixing and masses of the neutrinos. 5) The SM does not explain the origin and
naturalness of the hierarchy of mass scales in physics. 6) The SM does not accommodate cosmological inflation.
7) The SM does not include a quantum theory of gravity.

In the following, I discuss some of these issues, many of which will be addressed during Run~II of the LHC, and
most of which would be at least mitigated by supersymmetry, which remains my favourite framework for physics
beyond the SM.

\section{The End of the Universe may be Nigh}

\begin{figure}[htb]
\centering
\includegraphics[height=2.5in]{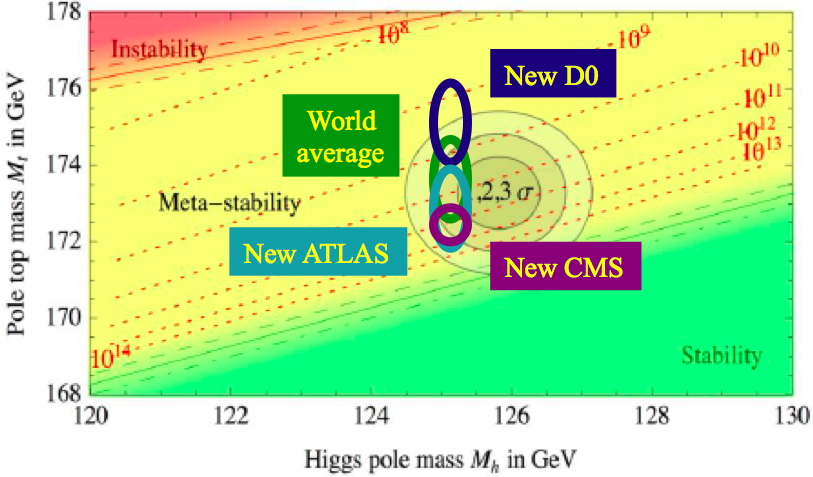}
\caption{\it The $(m_H, m_t)$ plane showing regions of vacuum (meta/in)stability~\cite{Buttazzo} and the 68\% CL
regions favoured by LHC measurements of $m_H$~\protect\cite{mH} and the 2014 world average
as well as more recent measurements of $m_t$~\protect\cite{WAmt}.}
\label{fig:mHmt}
\end{figure}

The quartic self-coupling $\lambda$ in the SM Higgs potential is renormalized by itself and,
more importantly, by its coupling to the top quark. This renormalization is negative, and tends to
drive $\lambda < 0$ at a scale $\Lambda_I$ estimated in~\cite{Buttazzo} to be
\begin{equation}
\log_{10} \left( \frac{\Lambda_I}{\rm GeV} \right) \; = \; 11.3 + 1.0 \left(\frac{m_H}{\rm GeV} - 126 \right)
- 1.2 \left( \frac{m_t}{\rm GeV} - 173.10 \right) + 0.4 \left( \frac{\alpha_s(m_Z) - 0.1184}{0.0007} \right) \, .
\label{LambdaI}
\end{equation}
Uisng the central values of the current world averages of $m_t$ (\ref{mt}), $m_H$ (\ref{mH}) and $\alpha_s$,
it seems that the current electroweak vacuum is metastable, as seen in Fig.~\ref{fig:mHmt}, and
adding the uncertainties in quadrature in (\ref{LambdaI}) (including $\Delta \alpha_s(m_Z) = \pm 0.0006$) one estimates
\begin{equation}
\log_{10} \left( \frac{\Lambda_I}{\rm GeV} \right) \; = \; 11.1 \pm 1.3 \, .
\label{LambdaIvalue}
\end{equation}
The new CMS measurement of $m_t$~\cite{CMSmt} would modify this to $\log_{10} (\Lambda_I/{\rm GeV}) = 11.6 \pm 0.7$
(the uncertainty would increase to $\pm 0.9$ if the uncertainty in $\alpha_s$ were doubled).
According to the SM, Higgs field values above this scale have a lower energy than the current
SM vacuum, which is therefore unstable. The estimate (\ref{LambdaIvalue}) will require re-evaluation
when the new CMS, D0 and ATLAS values of $m_t$ are incorporated in an updated world average,
but should one worry at all about this apparent instability? After all, the lifetime for vacuum decay is probably
much longer than the age of the Universe (just as well, otherwise we would not be here). However, I do consider this
to be a problem, for two reasons. One is that if the true vacuum is at some large Higgs value, how come our present
vacuum energy is very small in natural units? (Please do not throw an anthr*p*c argument at me!)
More cogently, if there is a lower-energy state out there, one would have expected (almost all of)
the Universe to have fallen into it because of the large quantum or thermal fluctuations in the hot and dense
early Universe~\cite{Hook}. (Again, please do not throw an anthr*p*c argument at me!) Averting this disaster requires
some new physics beyond the SM, maybe only dimension-6 operators or quantum gravity effects~\cite{Sher},
but supersymmetry would also do just fine~\cite{ER}.

\section{Supersymmetry: Dusk or Dawn?}

Supersymmetry still seems to me the most attractive scenario for physics beyond the SM, and I
would argue that Run~1 of the LHC has provided 3 additional motivations: supersymmetry would
stabilize our electroweak vacuum~\cite{ER}, simple supersymmetric models predicted correctly that the mass 
of the Higgs boson would be $\lesssim 130$~GeV~\cite{susymH}, and they also predicted correctly that its couplings
would be very similar to those in the SM~\cite{EHOW}. These are new arguments in addition to the familiar
arguments of naturalness, GUTs, superstrings and dark matter. Of course, we would be on safer
ground if supersymmetry had appeared during Run~1, but the conventional missing-energy 
searches have revealed nothing so far :( Should one be considering exotic signatures that might 
have been missed, or just push on hopefully to higher masses?

\begin{figure}[htb]
\centering
\includegraphics[height=2.5in]{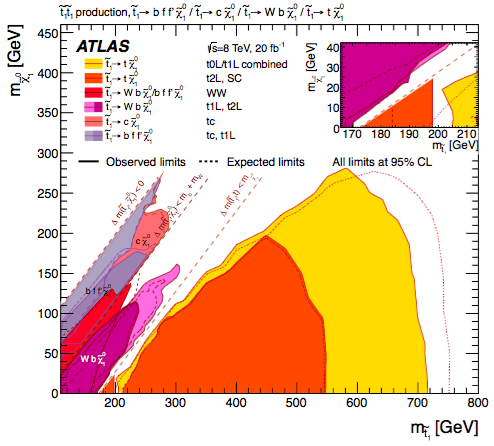}
\caption{\it The $(m_{\tilde t_1}, m_{\chi_1^0})$ plane showing the regions explored in various
searches during LHC Run~i~\protect\cite{Shchutska}, as well as the triangular region excluded by a measurement of the
${\bar t} t$ spin correlation~\protect\cite{Christinziani}.}
\label{fig:stops}
\end{figure}

The standard missing-energy signature is weakened in models with compressed
spectra, and ATLAS and CMS have been looking intensively at the possibility that the mass difference
between the lighter stop squark, ${\tilde t_1}$ and the lightest supersymmetric 
particle (LSP) $\chi_1^0$ might be small~\cite{Shchutska}, as in `natural supersymmetry' models~\cite{PRW}.
Depending how small, many different ${\tilde t_1}$ decays are potentially important, and it is possible that
the ${\tilde t_1}$ has fallen through the cracks in the searches, as seen in Fig.~\ref{fig:stops}. Interestingly,
calculations and measurements of the ${\bar t} t$ spin correlations are now sufficiently
precise to exclude one corner of this mosaic~\cite{Christinziani}.
Searches for ${\tilde t_1} \to t + \chi_1^0$ and ${\tilde t_1} \to b + $
chargino show promise for extending this search during Run~II~\cite{MC12}.

Another possibility is that the mass difference between the next-to-lightest sparticle (NLSP) and the LSP
may be so small that the LSP may show up in an LHC detector as a long-lived particle with a separated
decay vertex, or could even barrel out of the detector as a massive metastable charged particle~\cite{Shchutska}. Searches
for such signatures have often been interpreted in models where the NLSP is a chargino, but there are
also scenarios where it could be a charged slepton such as the lighter stau, ${\tilde \tau_1}$~\cite{MC12}.

Although these searches have not found anything yet, other supersymmetry-motivated searches have
found anomalies to whet the appetite~\cite{Shchutska}. For example, CMS has reported an `edge' feature in a dilepton
spectrum~\cite{CMSedge}, and ATLAS saw an excess of $Z$ events 
with missing transverse energy~\cite{ATLASZMET}. Let us see what happens
at Run~II. 

The same caution applies to other anomalies that have no reasonable supersymmetric interpretation, such
as the excesses of jet combinations with invariant masses around 2 and 5.2~TeV~\cite{Kersevan,CMS13,ATLAS13}.

\section{Energy Frontier vs Intensity Frontier}

\begin{figure}[htb]
\centering
\includegraphics[height=2.5in]{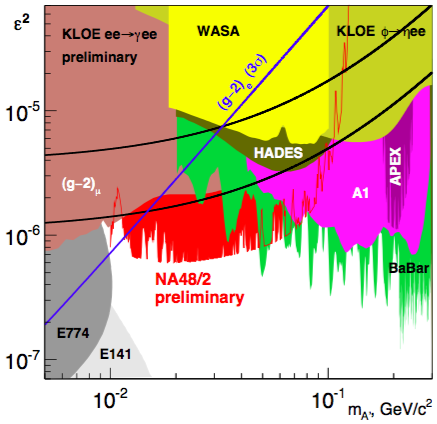}
\caption{\it The mass vs mixing plane for dark photon models, exhibiting the many experimental constraints,
which seem to exclude the region of interest for interpreting the $g_\mu - 2$ discrepancy~\protect\cite{Pospelov}.}
\label{fig:DarkPhoton}
\end{figure}

There are many examples of light, weakly-interacting particles that are difficult to detect at the LHC, but potentially
accessible to high-intensity lower-energy experiments~\cite{Pospelov} - though important constraints on 
some of these models are imposed by collider experiments such as BaBar and, most recently, LHCb. 
Examples include `dark photons', light dark matter particles, right-handed neutrinos and low-mass, 
weakly-coupled $Z^\prime$ bosons. Some of these models are of particular interest for explaining the 
discrepancy in $g_\mu - 2$. Unfortunately, as seen in Fig.~\ref{fig:DarkPhoton}, at least the dark photon
interpretation of $g_\mu - 2$ seems now to be excluded.

Projects for fixed-target or beam-dump experiments include HPS, DarkLight and BDX. A particularly
ambitious new project is SHiP~\cite{SHiP}, which would be able to explore many of these scenarios and also
have some bread-and-butter $\nu_\tau$ physics. There are also possibilities in the NA62 experiment that is designed 
to measure $K \to \pi \nu {\bar \nu}$, and ideas for installing a (low-energy) in an underground
laboratory and looking in a nearby large neutrino or proton decay detector for light, weakly-interacting
particles that it could produce~\cite{Pospelov}.

It is important to pursue such off-the-beaten-track ideas, particularly in view of the growing uncertainty
what new physics beyond the SM may be awaiting us at the energy frontier.

\section{Neutrino Physics}

This is the one area away from the energy frontier where new physics beyond the SM~\cite{Zhou} has been discovered, 
in the form of neutrino mixing and (presumably) masses, and experimental progress is being made rapidly.

In particular, great progress has been made with the
measurement of $\theta_{13}$ in the Daya Bay~\cite{DB}, RENO~\cite{RENO}
and Double Chooz~\cite{DC} reactor experiments~\cite{Stahl}.
The promising next step is the JUNO experiment~\cite{JUNO}, at a distance intermediate between KamLAND
and the current experiments, where both low-frequency `solar' and high-frequency `atmospheric'
neutrino oscillations will be important.

\begin{figure}[htb]
\centering
\includegraphics[height=2.5in]{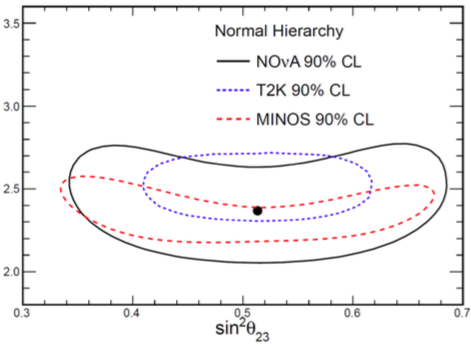}
\caption{\it First preliminary data from the NO$v$A accelerator neutrino experiment~\protect\cite{NOvA},
compared with previous data from the T2K~\protect\cite{T2K} and MINOS~\protect\cite{MINOS} experiments.}
\label{fig:NOvA}
\end{figure}

Progress on high-frequency oscillations was also reported at this Symposium~\cite{Shanahan}, with the observation of a fifth
$\nu_\mu \to \nu_\tau$ event by OPERA~\cite{OPERA}, amounting to a 5$\sigma$ discovery of this effect, and `first light'
from the NO$\nu$A experiment~\cite{NOvA}. With a fraction of the planned number of events, NO$\nu$A has already
obtained a constraint in the $(\sin^2 \theta_{23}, \Delta m^2)$ that is comparable with those from MINOS and T2K,
as seen in Fig.~\ref{fig:NOvA}.
The combination of T2K with reactor neutrino data already gave a mild preference for the normal hierarchy
of neutrino masses and for a non-zero CP-violating phase $\delta \sim 3 \pi/2$~\cite{T2K,Shanahan}, and both these
apparent trends are compatible with the NO$\nu$A data. We are a long way from a 5$\sigma$ discovery
of CP violation in neutrino oscillations, which will presumably require a new facility such as LBNF or
HyperKamiokande~\cite{Long}, but the augurs are promising.

Promising progress in the measurement of the absolute neutrino mass scale seems possible in the
(relatively) near future~\cite{Zavatarelli}. The grandiose Katrin experiment to measure the end-point of the 
Tritium $\beta$-decay spectrum
is scheduled to start taking data in 2016, and should have a 90\% CL upper limit sensitivity $\sim 0.2$~eV
after a full year of beam time. There are several projects to measure the neutrino mass via $e^-$ capture on
$^{163}$Ho. In parallel, several experiments are pursuing a possible Majorana neutrino mass via
neutrinoless double-$\beta$ decay, and their combination has reached an upper limt $\sim 0.13$~eV 
at the 90\% CL. Extensions of such experiments should eventually be able to reach into the range allowed by
models with an inverted hierarchy of neutrino masses, but the range allowed with the normal mass
hierarchy would be a challenge~\cite{Zavatarelli}. Meanwhile, the Planck measurements of the CMB have been used to
set an upper limit on the sum of neutrino masses of $\sim 0.2$~eV at the 68\% CL, and a possible future
CMB mission, the Cosmic Origins Explorer (COrE) could reach a sensitivity $< 60$~meV and measure the
absolute value of the sum of neutrino masses~\cite{Delabrouille}.

\section{Novel Ideas}

In this Section I describe 3 theoretical ideas that are novel: the latter two are very new, but the first has been around
for a decade, but has only recently been the object of a surge of interest.

\subsection{Twin Higgs and Neutral Naturalness}

The naturalness problem arises in the SM because quantum corrections to the squared mass of the Higgs boson
are out of control~\cite{McCullough}: in particular, there is a net quadratic divergence dominated by the heavy top quark, which is
cancelled in a supersymmetric theory by the stop squarks. The twin Higgs idea~\cite{twinH} is to introduce a different symmetry,
again by doubling the number of degrees of freedom, but this time with particles that are neutral under the SM
gauge group: instead, they form a copy of the SM with isomorphic but distinct couplings. The two copies $A, B$
are then connected via an SU(4) symmetry of the Higgs potential:
\begin{equation}
V \; \ni \; \lambda (|H_A|^2 + |H_B|^2) \, .
\label{twin}
\end{equation}
The idea is then to cancel the SM quadratic divergences with loops of SM-neutral particles: ``neutral naturalness''.
If one then postulates a negative squared mass for $H_B$, the underlying SU(4) symmetry is broken down to
SU(3), with 3 true Nambu-Goldstone bosons that are `eaten' by the $W^\pm_B$ and $Z_B$ and 4 PNGBs that
constitute `our' Higgs doublet. This model serves to broaden one's mind about possible LHC signals,
suggesting that the Higgs boson could decay into $B$-sector glueballs with displaced vertices~\cite{balls}.

\subsection{The Relaxion}

\begin{figure}[htb]
\centering
\includegraphics[height=2.5in]{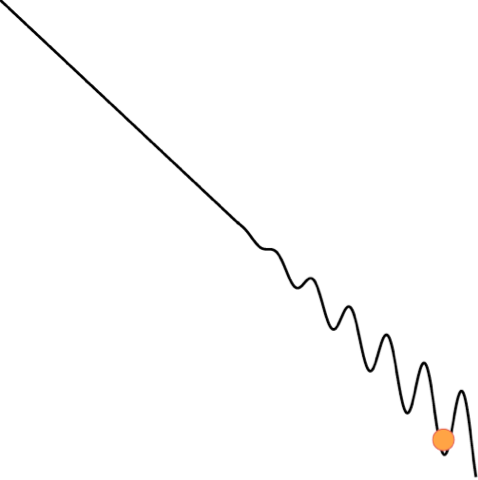}
\caption{\it The effective potential for the relaxion $\phi$~\protect\cite{Kaplan}.
After sliding down the linear part of the potential,
the relaxion is trapped in a local minimum once chiral symmetry is broken.}
\label{fig:relaxion}
\end{figure}

This suggestion~\cite{Kaplan} is to add to the SM an axion-like field $\phi$ with quite specific couplings:
\begin{equation}
{\cal L} \; = \; (M^2 - g \phi) |H|^2 - g M^2 \phi + \frac{\phi}{32 \pi^2 f} {\tilde G} G \, .
\label{relaxion1}
\end{equation}
After chiral symmetry breaking, the last term in (\ref{relaxion1}) is converted to the quark mass-dependent form
\begin{equation}
{\cal L}_{\chi SB} \; = \; f_\pi^3 m_q \cos \left( \frac{\phi}{f} \right) \; = \; f_\pi^3 \lambda_q \langle h \rangle \cos \left( \frac{\phi}{f} \right) \, ,
\label{relaxion2}
\end{equation}
where we have indicated the consequent implicit dependence of ${\cal L}_{\chi SB}$ on the Higgs vev.
One can now follow the cosmological evolution of the axion-like field $\phi$. It starts by sliding down the linear term
in the effective potential until $g \langle \phi \rangle > M^2$, at which point the Higgs develops a vev. When this happens,
the periodic ${\cal L}_{\chi SB}$ term (\ref{relaxion2}) switches on and grows with $\langle h \rangle$, trapping both
$\phi$ and the Higgs field, as seen in Fig.~\ref{fig:relaxion}. 
In particular, the Higgs vev $\langle h \rangle$ is naturally small, without a small parameter having been
introduced into the original Lagrangian (\ref{relaxion1}). However, this minimal relaxion model has some issues: it requires
some $10^{40}$ e-folds of inflation during the evolution, and the ground state has $\theta_{QCD} \ne 0$. The model
therefore needs epicycles, such as a new QCD-like sector~\cite{EGS}. In itself, however, the idea is very cute, and may have
applications elsewhere, e.g., in supersymmetric models~\cite{GnotY}.

\subsection{Composite Gauge Bosons}

It had long been thought that composite gauge bosons were impossible~\cite{WW,Komargodski}, but then along came an explicit
supersymmetric example with and SU(4) gauge group and 6 ($\mathbf{4} + \mathbf{\bar 4}$) multiplet pairs,
which yielded an SU(2) composite gauge theory with 12 $\mathbf{2}$ fermions and 32 singlet scalars in the
infrared limit. This behaviour was initially a puzzle, although consistent with the 4-dimensional $c$-theorem.
However, it is now known to be just one example of a strong-weak duality whose proof involves an abstruse
relationship between elliptic hypergeometric Gamma functions and q-Pochhammer symbols (!) that the
mathematicians have only recently discovered. It has been suggested that some deformation of this construction
might be applicable to the $\rho$ meson of QCD~\cite{Komargodski} - it has long been known that vector meson dominance
requires an effective dynamical local symmetry. Or perhaps this construction would be interesting in
dynamical models of electroweak symmetry breaking? Might the gauge bosons of the SM actually be
composite?

\section{Cosmology}

We have already discussed how CMB measurements can provide important constraints
on the neutrino mass scale, and the CMB also constrains the number of neutrino species~\cite{Delabrouille}.
In addition, the spectrum of fluctuations in the CMB provides
important information on the total density of the Universe (via the multipole moment
at which the first peak appears), requiring the existence of dark energy~\cite{Estrada},
and the densities of ordinary and dark matter
(via the relative heights of the other peaks in the spectrum).

\begin{figure}[htb]
\centering
\includegraphics[height=2.5in]{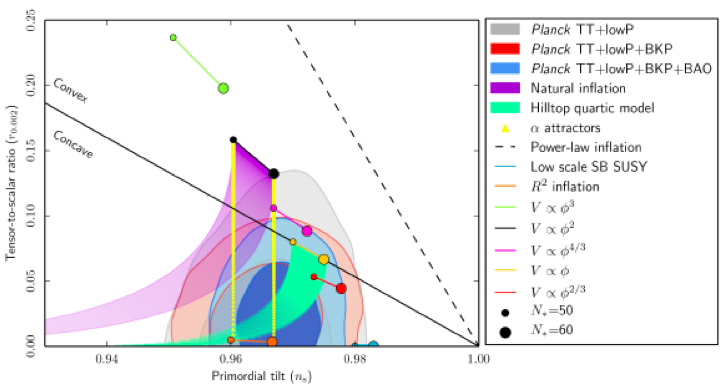}
\caption{\it The measurements of the tilt in the scalar CMB perturbations, $n_s$, and the
tensor-to-scalar ratio, $r$, constrain models of inflation~\protect\cite{Guth}. In particular, simple power-law
potentials are disfavoured~\protect\cite{Planck}, whereas the Starobinsky 
$R^2$ model~\protect\cite{Starobinsky} is highly consistent
with the data, as is Higgs inflation~\protect\cite{BezSh} and models based on no-scale supergravity~\protect\cite{ENO6}.}
\label{fig:CMB}
\end{figure}

These informative CMB fluctuations are thought to be due to an early inflationary
epoch~\cite{Guth}, and provide a window into possible physics at an energy scale far beyond any
conceivable accelerator. The data from the Planck satellite and other experiments on the
tilt in the spectrum of scalar perturbations, $n_s$ and the tensor-to-scalar ratio $r$
provide important constraints on models of inflation, excluding many simple models based
on simple powers of a single inflaton field, as seen in 
Fig.~\ref{fig:CMB}~\cite{Planck}. However, a number of models survive, notably
including the Starobinsky model that is based on a generalization of the Einstein action for
General Relativity with an extra $R^2$ term~\cite{Starobinsky}. Very similar predictions would be made by models
of Higgs inflation~\cite{BezSh} if they had a positive potential at large scales. However, as discussed
earlier, the effective potential of the SM seems likely to turn negative at some intermediate
scale, requiring some epicycle in the original minimal model of Higgs inflation, such as
supersymmetry~\cite{EHX}. Inflationary models based on no-scale supergravity can also make predictions
indistinguishable from those of the $R^2$ model~\cite{ENO6}. Interestingly,as seen in Fig.~\ref{fig:CMB},
the experimental limits on $n_s$ and $r$
begin to provide constraints on the number of e-folds during inflation, and hence the rate of
inflaton decays into lighter particles, which might provide another interesting connection between
cosmology and particle physics~\cite{EGNO5}.

\section{Gravitational Waves}

Nobody doubts the existence of gravitational waves, particularly since binary pulsars provide
strong indirect evidence for them~\cite{Barausse}. A non-zero value of $r$ would be evidence for gravitational
waves of quantum origin in the CMB, but the BICEP2 hopes have turned to dust~\cite{BICEPPlanck}. There are
several laser spectrometers such as LIGO, Virgo, GEO600 and KAGRA, and advanced LIGO
may soon have the capability to detect gravitational waves from mergers of neutron stars and/or
black holes. Another way to search for gravitational waves is via pulsar timing, which is closing in
on predictions based on black hole mergers, which are also among the prime targets of the
future eLISA experiment. We may not have long to wait before gravitational waves are
discovered, and the new window of gravitational astronomy opens on the Universe!

\section{Dark Matter}

There are many candidates for the dark matter, whose masses and coupling cover a large range.
Here I just focus on the fashionable WIMP scenario,as exemplified by the LSP. If WIMP
self-annihilations bring its density into the range allowed by cosmology, the annihilation
cross-section should be $\sim 3 \times 10^{-26}$~cm$^2$~\cite{Volansky}. However, in some models such as
supersymmetry, coannihilations with other particles that are almost degenerate could also be
important, in which case the self-annihilation cross-section could be smaller.

\begin{figure}[htb]
\centering
\includegraphics[height=2.5in]{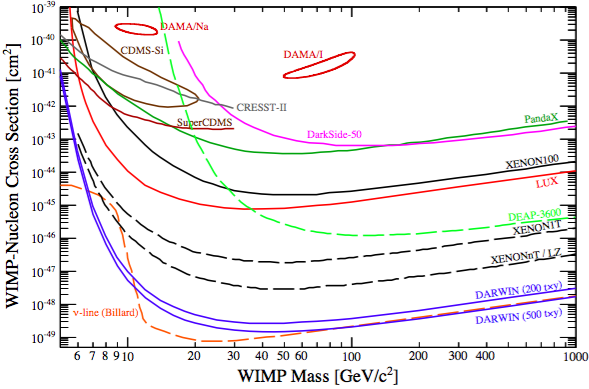}
\caption{\it A compilation of present limits and prospective future sensitivities to spin-independent
dark matter scattering~\protect\cite{Baudis}. The dashed orange line is the `neutrino floor'.}
\label{fig:Baudis}
\end{figure}

Fig.~\ref{fig:Baudis}
compiles the limits from many searches for the spin-independent direct scattering of WIMPs on
nuclei~\cite{Baudis}. Also shown is the `neutrino floor', below which there is a background from astrophysical
neutrinos. The present limits on the scattering cross-section are still some 3 orders of magnitude 
above this floor, but there are projects with sensitivities approaching it.

It is interesting to compare the direct searches for dark matter with those at the LHC using the
monojet signature~\cite{Malik}. In the cases of WIMPs that scatter spin-independently, the direct searches 
are more sensitive unless the WIMP mass $\lesssim 10$~GeV. However, the LHC monojet
searches are much more competitive in the spin-dependent case. Caution is in order, however,
since these comparisons are somewhat model-dependent, being sensitive to the mass and couplings of
whatever particle mediates the dark matter interaction with conventional matter.

\begin{figure}[htb]
\centering
\includegraphics[height=2.5in]{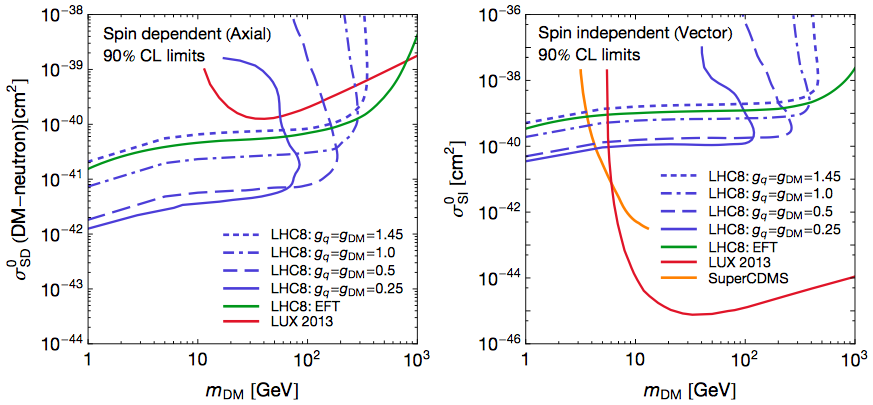}
\caption{\it A comparison of the sensitivities to spin-dependent (left panel) and spin-independent
(right panel) scattering of dark matter particles with LHC searches for monojets~\protect\cite{Malik}.}
\label{fig:Comparison}
\end{figure}

There have been many claims of indirect signals for the annihilations of astrophysical dark
matter particles. For example, there is a well-established excess of $\gamma$-rays from the
neighbourhood of the galactic centre in the GeV energy range~\cite{Berge}. However, there is evidence
that this excess could be due to unresolved point astrophysical sources~\cite{Slatyer}. The Fermi-LAT
Collaboration has also looked for $\gamma$-rays from dark matter annihilations~\cite{Sanchez}, notably
by combining data from dwarf galaxies, establishing an upper limit on the cross-section for WIMP
annihilations into $\tau^+ \tau^-$ that is below $3 \times 10^{-26}$~cm$^2$ for
$m_{\rm WIMP} \lesssim 70$~GeV~\cite{Berge}. There are also interesting limits dark matter self-annihilations
from the CMB.

Other options for dark matter searches include looking for positrons and antiprotons
produced by annihilations in the galactic halo. The AMS-02 experiment has confirmed
previous evidence for a large positron-to-electron ratio in the cosmic rays, but this could
be due to a nearby astrophysical source. The same experiment has recently reported a
high-statistics measurement of the antiproton-to-proton ratio, which is higher than
estimated in previous calculations assuming primary matter cosmic rays~\cite{Li}, but consistent
with more recent estimates of the cosmic-ray background~\cite{Annecy}.

On the other hand, the primary interest in looking for high-energy astrophysical neutrinos now seems to be to
search for astrophysical sources~\cite{vanE}.

\section{Future LHC Runs}

\begin{figure}[htb]
\centering
\includegraphics[height=2.5in]{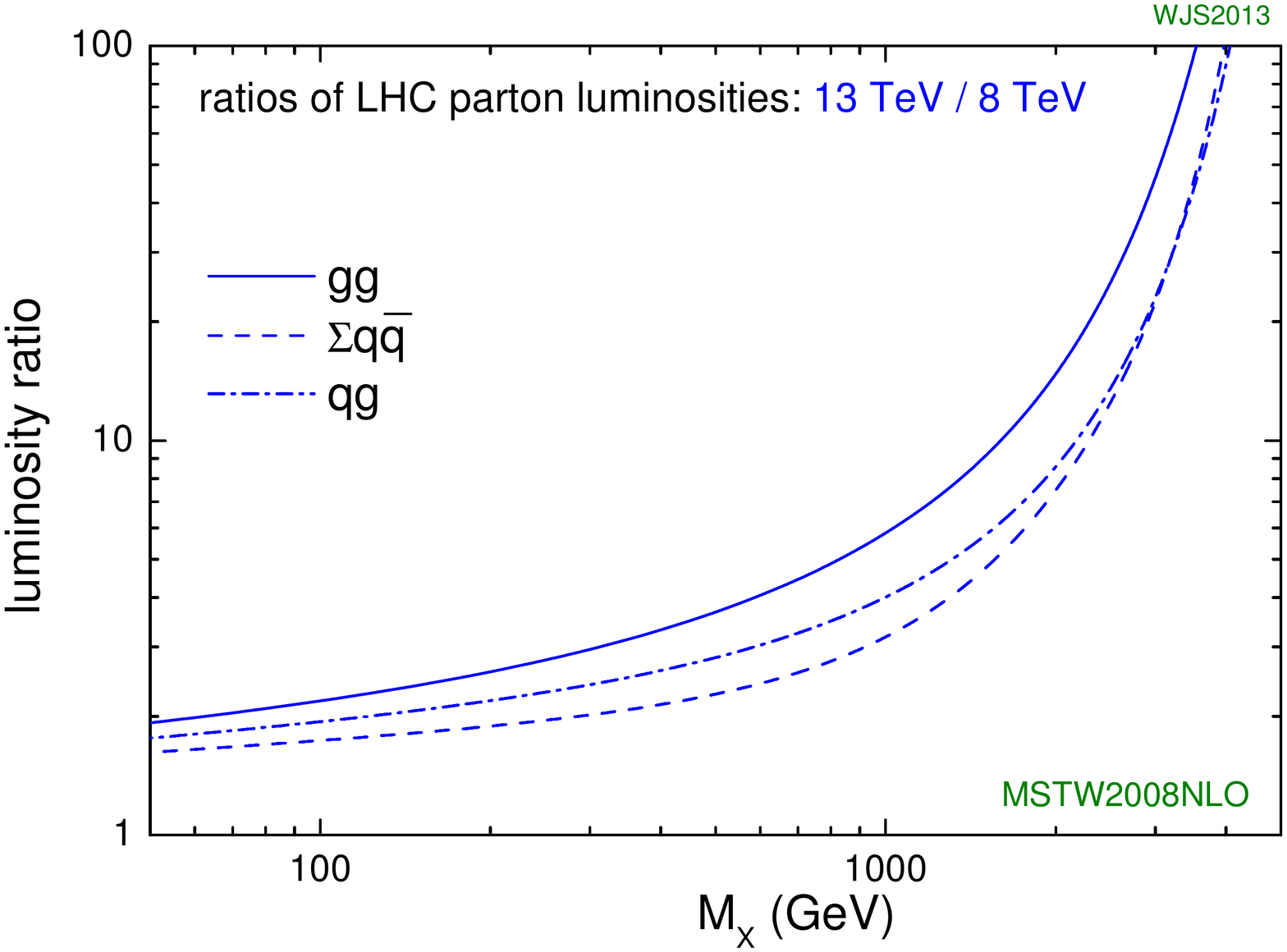}
\caption{\it The ratios of various parton-parton luminosities at 13 and 8~TeV, as functions of the
subprocess invariant masses~\protect\cite{CMS13,ATLAS13}.}
\label{fig:13over8}
\end{figure}

LHC Run~II at 13~TeV is underway. It is worth remembering the enormous amount of work
done during the long shutdown, and one must be philosophical about the inevitable
frustrations with SEUs in the QPS, TDIs, UFOs, a ULO and earth faults. Luminosity is
being accumulated, with ${\cal O}(3)$/fb expected by the end of 2015~\cite{Lamont}.
This is exciting because, for some types of heavy new physics weighing over 2~TeV, 1/fb at 13~TeV already
provides more reach than the 20/fb accumulated at 8/fb in Run~I, as seen in Fig.~\ref{fig:13over8}.

ATLAS~\cite{ATLAS13} and CMS~\cite{CMS13} have presented their first Run~II measurements, including
measurements of the total cross-sections for $W^\pm$, $Z^0$ and ${\bar t} t$
production. So far, all these early measurements are consistent with SM predictions.
However, each experiment has a showcase dijet event with an invariant mass $\sim 5.2$~TeV
to titillate the community.

The LHC is expected to accumulate $\sim 100$/fb in Run~I, and reach 300/fb
by the end of Run~II, after which the high-luminosity LHC should reach a total of 3000/fb.
So the LHC adventure has barely begun, with the energy already increased by a big
factor and more than two orders of magnitude of luminosity being planned. In addition
to pushing much further the direct searches for new physics beyond the SM, the future
LHC runs will provide much more detailed measurements of Higgs properties. Together,
these major elements of the future LHC programme will set the stage for possible
future collider projects.

\section{Future Accelerators}

The next collider to come online~\cite{det} will be the Super KEK-B, which will soon take many flavour
measurements to a new level, complementary to LHCb~\cite{Sakai}.

Looking further ahead, there is general interest in a higher-energy $e^+ e^-$ collider.
the most mature project is the ILC proposed for Japan~\cite{Komamiya}, aiming initially at a 
centre-of-mass energy of 500~GeV and eventually 1000~GeV. One alternative is the 
CLIC project being developed by a collaboration including CERN, which is aiming at 3~TeV in the
centre of mass~\cite{Roloff}, but is a less mature project. Recently interest has grown in a possible
circular $e^+ e^-$ collider, which would be limited to a centre-of-mass energy 
$\lesssim 350$~GeV but could achieve a higher luminosity than the current linear
collider designs~\cite{Benedikt}. One project is the CEPC proposed in China, which would have a
circumference of 50 to 70~km~\cite{CEPC}. Another project is the FCC-ee collider being studied
as a possible future CERN project~\cite{FCC-ee}, which would be located in a tunnel of 80 to 100~km
around Geneva, and aims at an even higher luminosity using technologies being deployed at
Super KEK-B.

\begin{figure}[htb]
\centering
\includegraphics[height=2.5in]{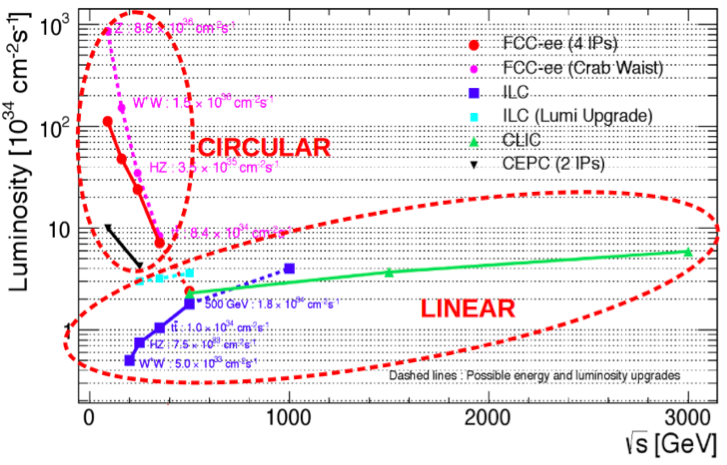}
\caption{\it A comparison of the possible centre-of-mass energies of proposed linear and circular $e^+ e^-$
colliders.}
\label{fig:Luminosities}
\end{figure}

Should the worldwide community prioritize energy or greater luminosity at lower energies?
Run~II of the LHC will provide crucial input. If the LHC discovers some new physics
beyond the SM, priority would go to studying it in more detail. If this new
physics includes particles light enough to be produced at an $e^+ e^-$ collider, that 
would surely be a top priority. If no such physics appears, studying the Higgs, $W^\pm$
and $Z^0$ in the greatest possible detail would assume a higher priority.

A large circular tunnel would also open up the possibility of a future higher-energy
$pp$ collider. For example, 16-Tesla magnets in a 100-km circumference tunnel,
as in the FCC-hh project now under study, would
enable a centre-of-mass energy of 100 TeV to be reached~\cite{Benedikt}. With an integrated
luminosity target of 20/ab~\cite{Mangano}, such a collider would provide enormous numbers of
Higgs bosons, and would have unequalled reach for new heavy particles. In combination,
such a high-energy $pp$ collider and an $e^+ e^-$ collider present a compelling
vision for the long-term future of high-energy physics, making possible both the
direct and indirect exploration of the 10~TeV scale.

Realizing such a vision will require co-operation among the different regions, engaging
deciders and the general public, and welcoming new communities and generations~\cite{Shaw}.

\end{document}